\documentclass[pdflatex,sn-mathphys-num] {sn-jnl}% Math and Physical Sciences Numbered Reference Style
\usepackage{graphicx}%
\usepackage{multirow}%
\usepackage{amsmath,amssymb,amsfonts}%
\usepackage{amsthm}%
\usepackage{mathrsfs}%
\usepackage[title]{appendix}%
\usepackage{xcolor}%
\usepackage{textcomp}%
\usepackage{manyfoot}%
\usepackage{booktabs}%
\usepackage{algorithm}%
\usepackage{algpseudocode}%
\usepackage{listings}%
\usepackage{dcolumn}% Align table columns on decimal point
\usepackage{bm}% bold math
\theoremstyle{thmstyleone}%
\theoremstyle{thmstyletwo}%

\theoremstyle{thmstylethree}%

\DeclareUnicodeCharacter{2217}{\textasteriskcentered}

\begin{document}

\title[Article Title]{HARLI CQUINN: Higher Adjusted Randomness with Linear In Complexity QUantum INspired Networks for K-Means}

%\title[Article Title]{ Optimization strategies using entanglement and data statistics in realizing Lloyd's quantum speed-up for quantum k-means algorithm}

%%=============================================================%%
%% GivenName	-> \fnm{Joergen W.}
%% Particle	-> \spfx{van der} -> surname prefix
%% FamilyName	-> \sur{Ploeg}
%% Suffix	-> \sfx{IV}
%% \author*[1,2]{\fnm{Joergen W.} \spfx{van der} \sur{Ploeg} 
%%  \sfx{IV}}\email{iauthor@gmail.com}
%%=============================================================%%

%\author*[1,2]{\fnm{First} \sur{Author}}\email{iauthor@gmail.com}

\author[1]{Jiten Oswal}\email{jiten.p.oswal@gmail.com}
\equalcont{}

\author[2]{Saumya Biswas}\email{sbiswas4@umd.edu}
\equalcont{These authors contributed equally to this work.}

\affil[1]{Independent Researcher, San Francisco, California, United States}

\affil[2]{Independent Researcher, College Park, Maryland, United States}

%\affil[1]{\orgname{Independent Researcher}, \orgaddress{\city{San Francisco}, \state{California}, \country{United States}}}

%\affil[2]{\orgname{Independent Researcher}, \orgaddress{\city{College Park}, \state{Maryland}, \country{United States}}}
%%==================================%%
%% Sample for unstructured abstract %%
%%==================================%%

\abstract{We contrast a minimalistic implementation of quantum k-means algorithm to classical k-means algorithm. With classical simulation results, we demonstrate a quantum performance, on and above par, with the classical k-means algorithm. We present benchmarks of its accuracy for test cases of both well-known and experimental datasets. Despite extensive research into quantum k-means algorithms, our approach reveals previously unexplored methodological improvements. The encoding step can be minimalistic with classical data imported into quantum states more directly than existing approaches. The proposed quantum-inspired algorithm performs better in terms of accuracy and Adjusted Rand Index (ARI) with respect to the bare classical k-means algorithm. By investigating multiple encoding strategies, we provide nuanced insights into quantum computational clustering techniques.}


%%\pacs[JEL Classification]{D8, H51}

%%\pacs[MSC Classification]{35A01, 65L10, 65L12, 65L20, 65L70}

\maketitle

\section{Introduction}\label{sec1}
k-means algorithm (aka Lloyd's algorithm) is an unsupervised Machine Learning (ML) method for compartmentalizing unlabeled datapoints into clusters of nearby datapoints \cite{lloyd1982least,chen2025provably}. Clustering, is a fundamental problem in ML and computer vision, especially in image classification \cite{caron2018deep}, segmentation \cite{coleman1979image}, tracking \cite{keuper2018motion}, network training \cite{coates2012learning} etc. It has garnered special attention as dual test-case for classical and ``quantized'' algorithms \cite{aimeur2007quantum}. A host of quantum or quantum inspired methods have been developed that improve the performance of the classical k-means algorithm \cite{kerenidis2019q,kerenidis2021quantum,doriguello2023you,lloyd2013quantum,viladomat2023quantum,qu2022performance,shao2023quantum,shi2020quantum,quiroga2021discriminating,bitsakos2024quantuml,ramirez2024advanced,banouar2024pattern,sarma2019quantum,bonny2020emulation,chen2025provably,gong2024quantum,zardini2024quantum,bui2025nisq} (in conjunction with genetic algorithms in \cite{xiao2008quantum,xiao2010quantum}, swarm intelligence algorithms in \cite{cui2022quantum,chen2020quantum,bai2021k}, fuzzy theory in \cite{hou2022quantum}, Quantum Approximate Optimization in \cite{saiphet2021quantum}, Quantum Spectral Clustering in \cite{li2022quantum}, continuous variable quantum circuits in \cite{haque2024continuous}, or quantum annealing in \cite{zaiou2021balanced,kumar2018quantum}). These propositions improve time or memory complexities. For n d-dimensional datapoints to be clustered into k clusters, the traditional k-means iteration has a time complexity of $\mathcal{O}(ndk)$ \cite{pedregosa2011scikit}. Refs. \cite{kerenidis2019q,doriguello2023you} improve the iteration time-costs to poly-logarithmic in n. Ref. \cite{aimeur2013quantum} proposed a quantum algorithm for finding k-median (a related problem) with complexity time $\mathcal{O}(n^{3/2}/\sqrt{k})$. On an economic number of qubits, even in classical simulations, the quantum inspired algorithms can demonstrate superior performance. In the NISQ (Noisy Intermediate Scale Quantum) era, quantum algorithms that may deliver on current hardware are of paramount importance \cite{khan2019k,bui2025nisq}. Quantum Information Science (QIS) has taken up the challenge of analyzing and optimizing such protocols in algorithmic frameworks called Quantum Machine Learning (QML).

Quantum clustering chooses a paradigm between Adiabatic Quantum Computing (AQC)\cite{zaech2024probabilistic,arthur2021balanced} or circuit-based quantum computation \cite{horn2001method,horn2001algorithm,wu2022quantum,ohno2022quantum}. In circuit based models, classical data is embedded into quantum circuits and quantum measurements can extricate distance estimates that can outperform classical Euclidean distance \cite{leporini2022efficient,ohno2022quantum}. It is customary to divide the subprocesses into dedicated quantum circuits \cite{poggiali2024quantum,khan2019k,chen2025provably}: subcircuit for distance measurement, subcircuit for cluster assignment etc. However, as an unsupervised approach, quantum k-means circuits serve a fundamentally different purpose than supervised classification approaches \cite{hur2022quantum}. Quantum circuits in k-means approach serves more to utilize quantum interference effects \cite{khan2019k}. Some works use quantum amplitude amplification approaches (based on e.g. Grover's algorithm)
\cite{aimeur2007quantum,gong2021quantum} leading to cryptographic and cybersecurity applications \cite{chen2020quantum,el2024quantum}. Far more ambitious proposals include speech encryption \cite{khaleel2021novel}, intelligent payment supervision \cite{li2024quantum}, modern power plant management (classical k-means) \cite{miraftabzadeh2023k}, cancer transcription analysis \cite{bonny2020cancer}, medical/patient data modeling \cite{deshmukh2023explainable,deshmukh2023patient}, multiband image segmentation \cite{casper2012quantum}, energy grid classification \cite{diadamo2022practical}, for high energy physics applications \cite{pires2021digital}.

Quantum inspired algorithms have built a legacy of their own. Short of the promise of a computation performed on a quantum machine, inspirations from quantum mechanics can design and solve problems on a classical computer with unique performance gains. Notable and pioneering examples are Quantum Inspired Evolutionary Algorithm (QEA) (Refs. \cite{han2002quantum,zhang2011quantum}). Ref. \cite{han2002quantum}'s revelation of QEAs solved combinatorial optimization problems on a classical computer. Other such families of algorithms are Quantum Particle Swarm Optimization algorithm (QPSO), Quantum Annealing Algorithm (QAA), Quantum Neural Network (QNN), Quantum Bayesian Network (QBN), Quantum Wavelet Transform (QWT), Quantum Clustering Algorithm (QC), etc \cite{li2020quantum}. These algorithms adopt fundamental laws of nature and quantum mechanics to solve complex problems. Remarkably, even before the actual arrival of widely affordable Quantum Processors, their simulations on a classical computer (on a limited number of qubits) are also known to outperform traditional classical algorithms.

\begin{algorithm}[t]
\caption{Standard k-means algorithm (one iteration)}\label{alg:Standard_k-means}
\begin{algorithmic}
\Require Data-Points $ \vec{v}_1, \ldots, \vec{v}_n \in \mathbb{R}^d$, \\
initial centers $\vec{c}_1^0, \ldots, \vec{c}_k^0 \in \mathbb{R}^d$
\Ensure Updated centers $ \vec{c}_1, \ldots, \vec{c}_k \in \mathbb{R}^d$ \\
Updated centroid assignment $ C_1, \ldots, C_n$, $ C_j \in \{ 1,2,..,k\}$

\bf{Cluster Assignment Phase:}
\For{$j = 1$ to $n$}
$C_j = \operatorname{argmin}_{i' \in [k]} \left\|\vec{v}_j - \vec{c}_{i'}^0\right\|$\;
\EndFor

\bf{Centroid Update Phase:}
\For{$i = 1$ to $k$}
$C_i^{\#} = \{ j | C_j = i \}$\;
\If{$|C_i^{\#}| > 0$}
$\vec{c}_i = \frac{1}{|C_i^{\#}|} \sum_{j \in C_i^{\#}} \vec{v}_j$\;
\EndIf
\EndFor
\end{algorithmic}
\end{algorithm}

We do experiments on improving the quantum k-means based on different centroid initializations, pre-sorting input data, and different encoding schemes. Such works have been done on classical k-means with a view to optimizing performance before \cite{arthur2021balanced,na2010research}. Ref. \cite{celebi2013comparative} investigated different centroid initializations, ref. \cite{ghosh2013comparative} investigated improvements with presorting of data. The quantum mechanical strategy taps into feature properties that classical methods are incapable of. For a handful of real world datasets, we have performed simulations of quantum inspired algorithms with several encoding schemes. Although the results have varied above and below par, a quantum inspired algorithm with angle encoding (linear complexity in qubit number) has been found to consistently perform above (or on par) the classical k-means algorithm. We present the results in terms of a well known figure of merit in classification algorithms of adjusted random index.

\section{Classical k-means algorithm}

In the most standard form (algorithm \ref{alg:Standard_k-means}), centroid coordinates from the last iteration is updated to minimize the sum of distances to the constituent datapoints \cite{lloyd1982least,mcqueen1967some}. The objective of the algorithm is to continually update the sets $C_i^{\#}$--that lists the assignment of each data-vector $\vec{v}_j$ into one of the k clusters. In the critical step of cluster assignment, each feature vector $\vec{v}_j$ is assigned to the nearest cluster (by minimizing distance to the cluster center
$\left\|\vec{v}_j-\vec{c}_{i^{\prime}}^0\right\|$). For the classical k-means() algorithm, the Euclidean distance is used. However, for the quantum algorithms, there are several options to choose from. Quantum distance metrics need not be proportional to the Euclidean distance, but should have a positive correlation with it. Quantum distance metrics have usually been derived form some geometric meaning of the associated Clifford spaces (product of $SU(2)$-spaces). What is defined as a distance in classical method, turns into an idea of ``similarity'' in quantum kernel methods. Overlap between quantum states quantify their similarity and is experimentally measured in the quantum algorithms. Datapoints with most similar wave functions are clustered together. Since, the quantum mechanical encoding is not unique, there is the natural question of how different encoding methods will perform in terms of final results.

%Algorithm 1 (Standard $k$-means (one iteration)).
%Input: , initial centers $\vec{c}_1^0, \ldots, \vec{c}_k^0 \in \mathbb{R}^d$
%1: for $j \in[k]$ do
%2: $\quad C_j^0=\left\{i \in[n]: j=\operatorname{argmin}_{j^{\prime} \in[k]}\left\|\vec{v}_i-\vec{c}_{j^{\prime}}^0\right\|\right\}$
%3: $\quad \vec{c}_j=\frac{1}{\left|C_j^0\right|} \sum_{i \in C_j^0} \vec{v}_i$
%Output: Updated centers $\vec{c}_1, \ldots, \vec{c}_k \in \mathbb{R}^d$

\section{Quantum Embedding}
In literature, several methods exist for quantum embedding of classical data and measurement of quantum distance \cite{berti2024role}. In this work, we are interested in finding more minimalist implementations of these embedding in terms of quantum operations. In the quantum setting, while several alternatives in terms of choice of distance metrics have been investigated \cite{leporini2022efficient}, alternatives in terms of quantum operations are less investigated. With well developed and widely accepted simulation packages, we show that several minimization in the necessary quantum operations maybe done without any significant deterioration in the efficiency or accuracy. We investigate three different encodings and compare their performances for a number of well-known datasets. Besides finding the optimal encoding method, we find recipes for better performance from statistical properties of the dataset.

QML is garnering attention as an advantageous platform over classical ML, especially in terms of kernel based classifiers \cite{thudumu2025supervised,jager2023universal}. In quantum k-means problems too, such improvements have been reported. Ref. \cite{benlamine2019distance} reports several advantageous schemes of quantum-centric distance metric that outperform classical distance based approaches. They reported superior performances (logarithmic in time complexity) based on quantum distance metrics proposed in \cite{wiebe2014quantum}. In fact, different quantum schemes \cite{wiebe2014quantum},\cite{lloyd2013quantum} or \cite{anagolum2019quantum} lead to varied levels of success rates for clustering. There exists popular methods for quantum circuit realizations for measuring the fidelity or overlap. The two broadly defined approaches \cite{el2024quantum} would be the controlled-SWAP test based approach \cite{gambs2006machine,benlamine2020quantum,diadamo2022practical} and Quantum Kernel-based approach. In this work, we adopt the latter. 

In the Controlled SWAP (CSWAP) test based encoding circuits (initially described in \cite{buhrman2001quantum}), 
to calculate the distance between quantum state encodings of classical data-vectors of centroid $\vec{x}$ and $\vec{c}$. The measurement furnishes an estimate of the overlap $|\langle\psi(\vec{x}) \mid \psi(\vec{c})\rangle|$ between their quantum state encodings. To this end, a quantum circuit with an ancillary qubit and additional qubits encoding the quantum states of the corresponding datapoints is availed of. Initially, the ancilla qubit is prepared in the superposition state $\frac{1}{\sqrt{2}}(|0\rangle+|1\rangle)$ with the help of a Hadamard gate. Next, quantum state encodings $|\psi(\vec{x})\rangle$ and $|\psi(\vec{c})\rangle$ are prepared into separate sets of qubits. This complete quantum state can be written as $\frac{1}{\sqrt{2}}(|0\rangle \otimes|\psi(\vec{x})\rangle \otimes|\psi(\vec{c})\rangle+ |1\rangle \otimes|\psi(\vec{x})\rangle \otimes|\psi(\vec{c})\rangle)$. A CSWAP gate then, (the ancilla qubit being the control qubit), is applied to the qubits containing $|\psi(\vec{x})\rangle$ and $|\psi(\vec{c})\rangle$. This results in the state $\frac{1}{\sqrt{2}}(|0\rangle \otimes|\psi(\vec{x})\rangle \otimes|\psi(\vec{c})\rangle+|1\rangle \otimes|\psi(\vec{c})\rangle \otimes|\psi(\vec{x})\rangle)$. A final Hadamard gate transforms the ancilla qubit leading the wavefunction to: $\frac{1}{2}(|0\rangle \otimes(|\psi(\vec{x})\rangle \otimes|\psi(\vec{c})\rangle+|\psi(\vec{c})\rangle \otimes|\psi(\vec{x})\rangle)+|1\rangle \otimes (|\psi(\vec{x})\rangle \otimes|\psi(\vec{c})\rangle-|\psi(\vec{c})\rangle \otimes|\psi(\vec{x})\rangle))$. In the final form of the wavefunction, the probability of measuring the ancilla qubit in the state $|0\rangle$ is given by $P(|0\rangle)=\frac{1+|\langle\psi(\vec{x}) \mid \psi(\vec{c})\rangle|^2}{2}$. From sufficient number of shots of the measurement, the overlap $|\langle\psi(\vec{x}) \mid \psi(\vec{c})\rangle|$
is estimated. 

\subsection{Quantum Kernel-based Approach}

While the CSWAP-test based method is quite popular, the quantum Kernel based method might be a more straightforward method for estimating the required overlap (or fidelity). The quantum kernel based approach is inspired by quantum Support Vector Machine theories \cite{rebentrost2014quantum} (Kernels can be estimated with the cSWAP-test as well \cite{moradi2022error}). It is closely associated with the method of quantum interference circuits introduced in \cite{khan2019k}. The overlap between datapoints $\vec{x}$ and centroids $\vec{c}$ can also be considered a quantum kernel $K(\vec{x},\vec{c})= |\langle \psi(\vec{c})|\psi(\vec{x}) \rangle|^2 $. And the process of encoding the classical vectors into quantum states is, in fact, a unitary transformations. Encoding of data point $\vec{x}$ into a quantum state using
a unitary operation $U(\vec{x})$ is $|\psi(\vec{x})\rangle=U(\vec{x})|\vec{0}\rangle$. Similarly, the centroid $\vec{c}$ can be encoded with a unitary operation $|\psi(\vec{c})\rangle$ using $U(\vec{c})$.
Therefore, the overlap, or similarity, between these states may be computed by
applying the adjoint unitary operation $U^{\dagger}(\vec{c})$ to $|\psi(\vec{x})\rangle$. yielding:
$K(\vec{x},\vec{c})$ (in Eq. \ref{eq_Kernel}). Here, instead of the separate set of qubits required in the CSWAP-test method, the same set of qubits is initialized in $|\vec{0} \rangle$, then subsequently operated on by $U(\vec{x})$ and $U^{\dagger}(\vec{c})$ sequentially. In the end, the probability of measuring the state in $|\vec{0} \rangle$ is simply $|\langle \psi(\vec{c})|\psi(\vec{x}) \rangle|^2$. This is a more minimalistic approach, requiring half the qubits and also simpler quantum operations. For the choice of angle encoding, we have to apply just one set of transformations instead of two; further simplifying the process. 

Our preferred method offers a more streamlined overlap estimation:
\begin{eqnarray}
K(\vec{x},\vec{c}) = |\langle\psi(\vec{c})|\psi(\vec{x})\rangle|^2 = |\langle\vec{0}|U^\dagger(\vec{c})U(\vec{x})|\vec{0}\rangle|^2 \ \ \ \ \ \label{eq_Kernel}
\end{eqnarray}

%\subsection{Quantum Kernel embedding}
When a tuple (or array) of numbers is mapped to a set of qubits, and the quantum state of the set of qubits is entangled, and quantum distance measurements account for quantum correlations that are not representable in classical distance formula. The quantum mechanical representations give access to correlations measurements that may reveal inter-relations of data-points otherwise inaccessible in the classical domain. The prospect of tapping into quantum correlations depend heavily on the choice of quantum encoding.

\subsubsection{Angle encoding} 
The individual components of the data vector $\vec{v}$ are encoded into individual qubits, and the overlap is measured simply, according to the Born rule of quantum mechanics. Since, a SU(2)- representation is limited by the limits of polar and azimuthal angles of a Bloch sphere, the data-components have to be scaled accordingly to encode into a Bloch sphere. We chose the \emph{MinMaxScaler} to scale the datasets, where datapoints are scaled into the range $[0,1]$ with the formula,
\begin{eqnarray}
    x_{scaled}= \frac{x - x_{min}}{x_{max}-x_{min}} \ \ \ \ \ \ \ \ \label{eq_x_scaled}
\end{eqnarray}

The unitary gates are the $SU(2)$-representation of the rotations of qubits \cite{benlamine2019distance},
\begin{eqnarray}
    U(\theta, \alpha, \lambda) = \begin{bmatrix} \cos\frac{\theta}{2} & -e^{i\lambda} \sin\frac{\theta}{2} \\ e^{i\alpha}\sin\frac{\theta}{2} & e^{i\lambda+i\alpha} \cos\frac{\theta}{2} \end{bmatrix}
\end{eqnarray}
As the typical course of action in angle encoding, we use the $R_y(\theta)$-rotations to encode the datapoints into the polar angle $\theta$ of the qubits \cite{weijun2024quantum}.
\begin{eqnarray}
    R_y(\theta_i) = \begin{bmatrix} \cos\frac{\theta_i}{2} & -\sin\frac{\theta_i}{2} \\ \sin\frac{\theta_i}{2} & \cos\frac{\theta_i}{2} \end{bmatrix} \ \ \ \ \ \label{eq_Ry}
\end{eqnarray}
A value of $\theta_i = \pi$ will flip the qubit state from $|0\rangle$ to $|1\rangle$ and vice versa. This is the range we want to use. 

Now, with the choice of $U = R_y(\theta)$, $U^{\dagger}(\vec{c})U(\vec{x})$ can be compactified into one rotation. This is the advantage of minimization that angle encoding can offer. Let's consider $a_k \in \vec{v}_j$ and $b_k \in \vec{C}_i$ (in algorithm \ref{alg:Standard_k-means}) to have n components i.e. $\vec{v}_j = (a_1, a_2,...,a_n)$ and $\vec{C}_i = (b_1, b_2,...,b_n)$. For $a_k \in \vec{v}_j$ and $b_k \in \vec{C}_i$ (in algorithm \ref{alg:Standard_k-means}) scaled with \emph{MinMaxScaler}, the difference components $d_k = a_k - b_k $ will be ranged in $[-1,1]$ and we encode $d_k$ into the $\theta_i$ as \cite{diadamo2022practical},
\begin{eqnarray}
\theta_k = (d_k+1)\frac{\pi}{2}  \end{eqnarray}
In our algorithm, we allocate single qubits to single data-points. For a tuple of n-dimensional data $\vec{x}=\left(x_1,x_2,...,x_n \right)$, the quantum encoding $|\psi \rangle$ is realized with rotating every single qubit with the rotation angle equaling the data-point. An arbitrary single quantum gate may be represented by $U_3(\theta, \phi)$. We choose to encode classical data with $\theta$. Specifically, the quantum state-encoding of $\vec{x}$ is,
\begin{eqnarray}
    | \psi \rangle = \otimes_{i=1}^n R_y(\theta_i) |0 \rangle^{\otimes n} \ \ \ \ \ \ \ \ \ \ \ \label{eq_angle_enc}
\end{eqnarray}
The important aspect of angle encoding is the complete absence of two qubit entangling gates in the preparation of quantum states. In the amplitude encoding (following subsection), the state preparation necessarily requires usage of two qubit entangling gates.

\subsubsection{Amplitude Encoding}
Amplitude encoding, (originally developed in \cite{mottonen2004transformation}) prepares a quantum state of $\log_2(M)$ qubits for a sequence of M-classical numbers ($c_j$) where each number translates into the probability amplitude of a basis state.
\begin{eqnarray}
    |\psi_{AE} \rangle = \sum_{j=0}^{M-1} c_j | a^j_0, a^j_1,..., a^j_{\log_2(M)}  \rangle = \sum_{j=0}^{M-1} c_j | j \rangle, \ \ \ \ \ \label{eq_Amp_enc}
\end{eqnarray}
where $|j \rangle \equiv | a^j_0, a^j_1, ..., a^j_{\log_2(M)} \rangle$ are $\log_2(M)$-qubit basis states, conveniently chosen as the binary sequences \cite{moradi2022clinical}. In this case the $\log_2(M)$ qubits are intricately entangled and a full exploitation of the entanglement promises of superior performance of the clustering task.

\subsubsection{Hybrid Encoding}
We choose the kernel based approach since it is the minimalistic approach. The method requires unitary transformations for preparing the quantum encoded states corresponding to the classical data vectors. The purpose of encoding is to transform the initial unentangled $|\vec{0}\rangle$ into an entangled state where quantum correlations will track the inter-dependence of all components of the datavectors more closely. Entanglement is an observer (measurement basis) dependent quantity \cite{zanardi2004quantum} and through the measurement results, we have access to will limit the access to this entanglement. If the measurement basis are individual qubit's $|0\rangle,|1\rangle$ states, then angle encoding will not entangle the qubits but amplitude encoding will. On the other hand, the amplitude encoding gives us access to an n-qubit entangled state even if the qubits are measured in their collective n-qubit basis. Therefore, we may expect the quantum k-means method to succeed better with amplitude encoding. In fact, comparing results for different encodings may reveal the role of quantum correlations in quantum k-means algorithm.

As the unentangled encoding, the angle encoding is expected to produce the least accuracy. To establish the utility of quantum entanglement in the clustering algorithm, we define an encoding in between the angle encoding and amplitude encoding. It is defined as encoding two components of the datavector (of length M) into a qubit and the rest of (M-2) into (M-2) qubits in the manner of angle encoding (eq. \eqref{eq_angle_enc}). The encoding requires (M-1) qubits and helps us identify the two features that contribute the most to the classification task. If entanglement is in fact a useful resource, this hybrid encoding would outperform the angle encoding clustering method. Moreover, we investigate a systematic way of choosing these two numbers from M-numbers of datavectors that may maximize this accuracy. This flushes out the most significant data-components of the datavector as well as the statistical criterion that signify the contribution in a k-means clustering problem. We seek for this criterion by arranging the numbers in increasing or decreasing orders of spread, kurtosis, and skewness; and choosing the two leading numbers for amplitude encoding.

We finish this section by commenting on the entanglement properties of the traditional CSWAP-based encoding method. Given $M=2^n$ dimensional complex vectors $\vec{x}$ and $\vec{w}$ with components $x_j=\left|x_j\right| e^{-i \alpha_j}$ and $w_j=\left|w_j\right| e^{-i \beta_j}$ respectively, the quantum encoding in the method is \cite{benlamine2019distance,benlamine2020quantum,wiebe2014quantum},
\begin{eqnarray}
\begin{aligned}
& |\psi\rangle=\frac{1}{\sqrt{d}} \sum_j|j\rangle\left(\sqrt{1-\frac{\left|x_j\right|^2}{r_{\max }^2}} e^{-i \alpha_j}|0\rangle+\frac{x_j}{r_{\max }}|1\rangle\right)|1\rangle \\
& |\phi\rangle=\frac{1}{\sqrt{d}} \sum_j|j\rangle|1\rangle\left(\sqrt{1-\frac{\left|w_j\right|^2}{r_{\max }^2}} e^{-i \beta_j}|0\rangle+\frac{w_j}{r_{\max }}|1\rangle\right),
\end{aligned}
\end{eqnarray}

where $j=\{1, \ldots, n\}$, and $r_{\text {max }}$ is an upper bound for any feature in the dataset (possibly from standardization via scaling). The ancilla qubit is entangled with the qubit register in the realization of $\psi \rangle$ and $\phi \rangle$, and we expect high accuracies of an entangled register. In literature, such high performances have been reported.

\subsection{Quantum Distance metrics}
A Quantum distance metric does not need to be proportional to the classical Euclidean distance, but should have a positive correlƒformation with it \cite{benlamine2019distance}. The common hypothesis is that distance metrics can tap into the geometrical features of the quantum mechanical states \cite{leporini2022efficient}, and possibly the data and hence serve as a distance measure. Distance measures in quantum information theory have served many important purposes such as quantifying entanglement, optimizing quantum control processes, and quantum error correction \cite{mendoncca2008alternative}. The set of state matrices or density matrices can be elevated to a Riemannian manifold where different metrics may be defined for QIS purposes. A commonly investigated quantity is fidelity which in itself is not a metric, but enable the definition of several metrics who may serve as quantum distance measures. Fidelity is a measure of similarity between quantum states and can be estimated from experiments quite conveniently. We take this route in our numerical simulations. We use the so-called ``Bures distance''-metric from an experimental estimate of Uhlmann-Jozsa fidelity $\mathcal{F}$. It is a distance measure between quantum mechanical states $\rho$ and $\sigma$ defined in terms of transition probabilities \cite{uhlmann1976transition,jozsa1994fidelity}.
\begin{eqnarray}
    \mathcal{F}(\rho, \sigma) = \max\limits_{\substack{|\psi\rangle, |\phi\rangle}} |\langle \psi|\phi \rangle|^2, \ \ \ \ \ \ \ \ \ \ \label{eq_UJ_fidelity}
\end{eqnarray}
where $|\psi\rangle, |\phi \rangle$-are restricted to be purifications of $\rho$ and $\sigma$. The Bures distance metric is defined, in turn, as \cite{hubner1992explicit,bures1969extension}
\begin{eqnarray}
    B\left[ \mathcal{F}(\rho, \sigma) \right] = \sqrt{2-2\sqrt{\mathcal{F}(\rho, \sigma)}}. \ \ \ \ \ \ \ \ \ \ \label{eq_Bures_distance}
\end{eqnarray}
While the classical ``Euclidean'' distance \cite{li2016novel,mo2016generalized} is still available in a quantum mechanical state representation, the so-called ``trace distance'' (Eq. \eqref{eq_tr}), the geometric distance metrics have shown superior performance in recent times \cite{benlamine2019distance,leporini2022efficient}.
\begin{eqnarray}
    \mathcal{D}(\rho, \sigma) = \frac{1}{2}||\rho- \sigma||_{tr} \ \ \ \ \ \ \ \label{eq_tr}
\end{eqnarray}

\subsection{\label{sec:quantum inspired methods}Quantum Inspired Algorithms}
There are potential benefits in QML that initial guesses can be procured from cleverly formulated superpositions that may evade the challenge of local minima. We use such an empirically inspired ansatz circuit to sample from diverse initial centroids. A quantum state need not be a single state, it can be a superposition of several. Therefore, it would be a wasted opportunity not to use a superposition of states. Such a scheme proceeds with parallel evolution of several evolutions of individual constituent states. A given initial state guess may be prone to a given minima. Several initial states can possibly overcome their individual pitfalls and reach the global minima for the problem. 

A sufficient number of experimental runs provides an estimate of state overlap. We adopt a quantum distance metric obtained by weighting the quantum state overlap with the classical Euclidean distance \cite{khan2019k},
\begin{eqnarray}
  d_{\text{quantum}} = 2 \times fidelity \times d_{\text{classical}} \ \ \ \label{eq_d_quantum_overlap}
\end{eqnarray}

%$$d_{\text{quantum}} = \frac{\text{total_ones}}{\text{total_bits}} \times d_{\text{classical}} \times 2$$

Our investigated quantum inspired algorithms have the same steps as the classical k-means algorithm described in the algorithm \ref{alg:Standard_k-means}, only enhanced by a quantum enhanced initialization step and alternate definition of a ``quantum distance''. This approach leverages quantum superposition to find an initial centroid guess from an educated guess based on a superposition state. First centroid is chosen randomly, and the rest from the usage of a specialized quantum circuit to sample from diverse initial centroids' sample-space. Controlled rotations based on distances to existing centroids are applied to find the remaining ones followed by a measurement to collapse the superposition. This approach helps avoid poor local minima by selecting well-separated initial centroids. We can exploit the algorithms on a classical computer with simulators. Although the runtimes are long and datasizes are restricted, a few real world datasets can be successfully investigated.

\subsection{Figures of merit}
In this work, we are interested in the success rate of clustering, but not the convergence properties. In fact, the convergence criterion is checked several times. Only if the criterion is satisfied a few times (typically 3--5; exact number determined from multiple runs) in a row, we terminate the iterations. If the convergence criterion is overturned in less than the required number of iterations (indicating a possible local minima), we reset the counting toward this required number and freshly continue the iterations.

The evaluation criteria we are interested in are Adjusted Random Index (ARI) and silhouette score (SS) \cite{zhu2021improved,arthur2021balanced}. Traditionally, the ARI is chosen when the true labels are known and we can fairly evaluate the efficiencies of the clustering algorithm with them. This is a measure of the actual agreement with the ground truth. On the other hand, SS is useful when the true labels are not available; it is a measure of cohesion and similarity of the constituents among the clusters that the algorithm produced and how they contrast with other cluster's constituents. The two definitions are discussed below.

For the true category information (ground truth) $C$ and the clustering result $K$, the number of pairs of elements that are both in the same category in $C$ and $K$ (correctly labeled by the algorithm) being a, and the number of pairs of elements that are in different categories in $C$ and $K$ (incorrectly labeled by the algorithm) being b, the so-called Random Index (RI) is defined as:

\begin{eqnarray}
R I=\frac{a+b}{C_2^{n_{\text {samples }}}}    \ \ \ \ \ \ \label{eq_RI}
\end{eqnarray}
where $C_2^{n_{\text {samples }}}$ is the total number of pairs of elements formed from the data set. The range of $R I$ is $[0,1]$. The closer the value of $R I$ is to 1, the better is the agreement with the true labels and the higher the accuracy of the algorithm. Unfortunately, a random division of the data does not provide a close to zero constant $R I$. As an improvement the Adjusted Random Index (ARI) is defined. It quantifies a higher degree of discrimination. The $A R I$ is defined as:

\begin{eqnarray}
A R I=\frac{R I-E[R I]}{\max [R I]-E[R I]}    \ \ \ \ \ \ \label{eq_ARI}
\end{eqnarray}
where $R I$ is the Rand index and $\mathrm{E}[R I]$ is the mean. $A R I$ has a range of $[-1,1]$. A larger value for $A R I$ indicates a higher success in detecting true cluster labels. A low value of $R I$ would indicate an inefficient clustering performance; especially a negative value would indicate a total failure.

The second figure of merit of interest to us is Silhouette Score for the entire data defined as the average of Silhouette Coefficients defined for individual samples. It embodies a measure of cohesion and separation among the samples after clustering. After a clustering has been performed, let $A$ be the average distance between the sample and the other samples in its cluster, and $B$ the average distance between the sample and the samples in the other clusters, the silhouette coefficient of the sample is defined as:
\begin{eqnarray}
    S=\frac{B-A}{\max (A, B)} \ \ \ \ \ \ \ \ \ \label{eq_SC}
\end{eqnarray}
where $S$ denotes the silhouette coefficient of a single sample. The Silhouette Score is the average value of all sample silhouette coefficients. It ranges between $-1$ and 1. A SS closer to 1 indicates more successful clustering performance, and conversely, SS closer to -1 indicates unsatisfactory clustering performance.

\section{Results}
The results reported here are limited to data scaling by MinMaxScaler only. For a fair comparison, we only present results for dataset with true labels known. This way, we can have a proper estimate of ARI and compare the performance of the algorithms equitably. For the results presented below, we attempt the classifications on the basis of reasonable target and features where a reasonable result is obtainable. When multiple target is available, we choose the target that produces the best results. For example, for the Palmer Penguins dataset, the clustering may be attempted based on Island, species, or sex. Since, we find the best ARI with Islands, we present the comparative performance based on this attribute.

\subsection{Iris Dataset}
The Iris dataset has four features and three clusters corresponding to the three species of iris (flowers). All three quantum algorithms succeed in clustering into the three categories better than the classical sklearn algorithm. Especially, the arrangement of features in the ascending order of Kurtosis yields a much higher value for ARI for the hybrid encoding.

\begin{table}[h]
\caption{Performance indicators of Clustering algorithms on the iris dataset.}\label{tab:iris}
\begin{tabular*}{\textwidth}{@{\extracolsep\fill}lcccc}
\toprule%
& \multicolumn{1}{@{}c@{}}{Classical k-means} & \multicolumn{3}{@{}c@{}}{Quantum k-means} \\\cmidrule{2-2}\cmidrule{3-5}%
Figure of merit & sklearn & Angle Encoding & Amplitude Encoding & Hybrid Encoding \\
\midrule
Silhouette Score  & 0.4829 & 0.4806 & 0.5022 & 0.4821\\
Adjusted Random Index  & 0.7009 & 0.7430 & 0.7028 & 0.8857\\
\botrule
\end{tabular*}
\end{table}

\subsection{Wine Dataset}
The Wine dataset presents the chemical analysis of three different cultivars of wine in terms of thirteen features. For this dataset, angle encoding works on par with the sklearn k-means; but amplitude and hybrid encoding somewhat underperforms to cluster into the three categories.

\begin{table}[h]
\caption{Performance indicators of Clustering algorithms on the wine dataset.}\label{tab:wine}
\begin{tabular*}{\textwidth}{@{\extracolsep\fill}lcccc}
\toprule%
& \multicolumn{1}{@{}c@{}}{Classical k-means} & \multicolumn{3}{@{}c@{}}{Quantum k-means} \\\cmidrule{2-2}\cmidrule{3-5}%
Figure of merit & sklearn & Angle Encoding & Amplitude Encoding & Hybrid Encoding \\
\midrule
Silhouette Score  & 0.3009 & 0.3013 & 0.2949 & 0.2986\\
Adjusted Random Index  & 0.8537 & 0.8685 & 0.8000 & 0.8040\\
\botrule
\end{tabular*}
\end{table}

\subsection{Seeds Dataset}
The three varieties of seeds are presented with the seven features. Quantum algorithms perform better in terms of the ARI and on par in terms of the SS with the sklearn algorithm.

\begin{table}[h]
\caption{Performance indicators of Clustering algorithms on the seeds dataset.}\label{tab:seeds}
\begin{tabular*}{\textwidth}{@{\extracolsep\fill}lcccc}
\toprule%
& \multicolumn{1}{@{}c@{}}{Classical k-means} & \multicolumn{3}{@{}c@{}}{Quantum k-means} \\\cmidrule{2-2}\cmidrule{3-5}%
Figure of merit & sklearn & Angle Encoding & Amplitude Encoding & Hybrid Encoding \\
\midrule
Silhouette Score  & 0.4221 & 0.4183 & 0.4141 & 0.4221\\
Adjusted Random Index  & 0.6934 & 0.7628 & 0.8109 & 0.7049\\
\botrule
\end{tabular*}
\end{table}

\subsection{glass Dataset}
Six types of glass with nine features is attempted to be categorized. All algorithms perform nearly the same for the dataset, both in terms of ARI and SS.

\begin{table}[h]
\caption{Performance indicators of Clustering algorithms on the glass dataset.}\label{tab2}
\begin{tabular*}{\textwidth}{@{\extracolsep\fill}lcccc}
\toprule%
& \multicolumn{1}{@{}c@{}}{Classical k-means} & \multicolumn{3}{@{}c@{}}{Quantum k-means} \\\cmidrule{2-2}\cmidrule{3-5}%
Figure of merit & sklearn & Angle Encoding & Amplitude Encoding & Hybrid Encoding \\
\midrule
Silhouette Score  & 0.3796 & 0.3673 & 0.3687 & 0.3635\\
Adjusted Random Index  & 0.1617 & 0.1603 & 0.1667 & 0.1685\\
\botrule
\end{tabular*}
\end{table}

\subsection{Palmer Penguins Dataset}
The Palmer Penguins dataset lists features of three species of Penguins from three islands. The dataset may be categorized based on island, species, or sex. We attempt the clustering on the basis of islands and five features. For this dataset as well, all algorithms have nearly matched performance.

\begin{table}[h]
\caption{Performance indicators of Clustering algorithms on the Palmer Penguins dataset.}\label{tab:penguins}
\begin{tabular*}{\textwidth}{@{\extracolsep\fill}lcccc}
\toprule%
& \multicolumn{1}{@{}c@{}}{Classical k-means} & \multicolumn{3}{@{}c@{}}{Quantum k-means} \\\cmidrule{2-2}\cmidrule{3-5}%
Figure of merit & sklearn & Angle Encoding & Amplitude Encoding & Hybrid Encoding \\
\midrule
Silhouette Score  & 0.4329 & 0.4222 & 0.4284 & 0.3871\\
Adjusted Random Index  & 0.3697 & 0.3771 & 0.3774 & 0.3797\\
\botrule
\end{tabular*}
\end{table}

\subsection{Algerian Forest Fire Dataset}
The dataset has data on two regions. We discard the data for Sidi-Bel Abbes Region data and study the data for Bejaia Region Dataset only. We perform this classification with ten features leaving out day, month, and year entries. The quantum algorithms marginally perform better in finding these six clusters.

\begin{table}[h]
\caption{Performance indicators of Clustering algorithms on the Algerian Forest Fires dataset.}\label{tab:Algerian Forest fires}
\begin{tabular*}{\textwidth}{@{\extracolsep\fill}lcccc}
\toprule%
& \multicolumn{1}{@{}c@{}}{Classical k-means} & \multicolumn{3}{@{}c@{}}{Quantum k-means} \\\cmidrule{2-2}\cmidrule{3-5}%
Figure of merit & sklearn & Angle Encoding & Amplitude Encoding & Hybrid Encoding \\
\midrule
Silhouette Score  & 0.2103 & 0.2191 & 0.2066 & 0.2048\\
Adjusted Random Index  & 0.1693 & 0.2617 & 0.2123 & 0.2014\\
\botrule
\end{tabular*}
\end{table}

\subsection{Wholesale customers Dataset}
For the wholesale customers dataset, we perform the clustering based on the two channels on six features. Interestingly, the classical sklearn algorithm fails (gives a negative ARI value). The quantum algorithms, on the other hand, can provide quite high values of ARI. Since, the angle and amplitude encoding can provide high SS values as well, the utility and efficacy of quantum algorithms are quite tangible for this dataset.

\begin{table}[h]
\caption{Performance indicators of Clustering algorithms on the wholesale customers dataset.}\label{tab:wholesale customers}
\begin{tabular*}{\textwidth}{@{\extracolsep\fill}lcccc}
\toprule%
& \multicolumn{1}{@{}c@{}}{Classical k-means} & \multicolumn{3}{@{}c@{}}{Quantum k-means} \\\cmidrule{2-2}\cmidrule{3-5}%
Figure of merit & sklearn & Angle Encoding & Amplitude Encoding & Hybrid Encoding \\
\midrule
Silhouette Score  & 0.2871 & 0.5385 & 0.3257 & 0.1119\\
Adjusted Random Index  & -0.0364(fail) & 0.3061 & 0.4627 & 0.1504\\
\botrule
\end{tabular*}
\end{table}

\subsection{E Coli Dataset}
The clustering is attempted in terms of protein localization sites. In the original dataset, out of the eight site labels, two (`imS' and `imL') appear for 2 times each. We get rid of these two categories and datavectors (4 rows of the 336 rows) and perform the clustering on the remaining six clusters.

\begin{table}[!h]
\caption{Performance indicators of Clustering algorithms on the E Coli dataset.}\label{tab:ecoli}
\begin{tabular*}{\textwidth}{@{\extracolsep\fill}lcccc}
\toprule%
& \multicolumn{1}{@{}c@{}}{Classical k-means} & \multicolumn{3}{@{}c@{}}{Quantum k-means} \\\cmidrule{2-2}\cmidrule{3-5}%
Figure of merit & sklearn & Angle Encoding & Amplitude Encoding & Hybrid Encoding \\
\midrule
Silhouette Score  & 0.2937 & 0.2540 & 0.2988 & 0.2478\\
Adjusted Random Index  & 0.4536 & 0.4745 & 0.5487 & 0.4501\\
\botrule
\end{tabular*}
\end{table}

\section{Discussions and Limitations}
The results reported here is limited to MinMaxScaler only; other scalers like MaxAbsScaler that also scale features to a finite range may be mapped to a quantum encoding as well. For uniformity, we have used the same scaler throughout. Scalers that do not restrict to a finite range have to be formulated differently since SU(2)-spheres have a finite range for angles.

The mapping of scaled feature values into qubit's angles introduce a curving to the features. Upon the mapping, the feature values now curiously get a probabilistic connotation from the Born rule of quantum mechanics. The probability of measurement in the all-zeros ($|00..0\rangle$) qubit state now comes from the composite probability dictated by the product of $\cos$ of all the angles. Interestingly, the geometric measure as the quantum distance has outperformed the classical sklearn function in all the investigation in terms of ARI. This result is quite promising, since ARI is a measure of absolute success compared against the ground truth. The downside of amplitude encoding is the expense of qubit resources. The resource complexity is linear in qubit number. Therefore for high dimensional data, angle encoding may not be a feasible prospect.

In the pure form, amplitude encoding leads to an overlap or fidelity estimation that is not different form the classical trace distance. Although, it is converted into the Bures distance or a geometrical measure subsequently, this encoding performs in a similar manner as the classical distance based method. While it is economic in terms of qubit resources (logarithmic in qubit number), it has underperformed for some of the datasets (e.g. the Iris-dataset). As a quantum inspired method to be used on classical machines, there may be further challenges since preparation of the amplitude encoded state is somehow computation intensive requiring high runtimes.

We investigated the hybrid method as a measure for interpolating the performance between the amplitude and angle encoding. While, angle encoding has consistently succeeded in clustering on and above par, we investigate the possibility of improving the resource requirements by one qubit as a step toward economizing. The choice of the feature values that should be amplitude encoded is dictated by the comparative statistical properties of the features. This problem requires further investigations. While we choose the two numbers by maximum Kurtosis among the features, much more datasets should be explored to validate this hypothesis. Overall, the hybrid encoding has decent performance. However, it may over- or under-perform the baseline classical k-means algorithm. Therefore, for a unknown dataset without true labels, we cannot ascertain an improved clustering in the formulated hybrid encoding. 

While all data from nature has inherent mathematical structure, there is a question of sufficient representation of the mathematical model in a given dataset. Artificially created datasets can provide such datasets with inherent mathematical pattern. For a validation of the performance, such artificial datasets may be useful. Even in the absence of known or tangible mathematical model, for the real  life datasets, angle encoding has consistently performed better than the classical k-means algorithm. The method should be investigated further as a promising improvement over the classical algorithm.

QML is believed to reap the benefits of Quantum parallelism. The quantum states can host an exponential in size classical numbers that can be tracked in parallel. This may enable clustering in higher-dimensional quantum feature spaces, potentially capturing more complex patterns. Since our current implementations run on classical simulators, we do not venture into these high dimensional data yet. Simulation overhead makes quantum approaches slower. On real quantum hardware, the algorithm could potentially realize a speedup for distance calculations through quantum parallelism and better handling of high-dimensional data.

Some points of academic interest are in novel distance metrics based on quantum state properties and quantum amplitude amplification. There are several geometrical distance measures that quantum inspired algorithms can exploit. In this work, we only use the Bures distance. Additionally, Quantum Amplitude Amplification (Grover's algorithm) could provide further speedup in searching for optimal centroids. This is a step we may use in future to seek further improvements.

\section{Conclusions and Outlook}
QML derives its power from tapping into quantum correlations that has no classical counterpart. In fact, in this work we found that quantum inspired or enhanced encoding can in fact outperform classical k-means algorithms by far for some datasets. However, this trait has not been found consistently across all datasets. The statistical and inherent properties of the dataset should be investigated further for understanding the interrelations. The holy grail in QML today is a machine learning problem and/or dataset where classical ML algorithms cannot perform a task as well as QML algorithms. A dataset or problem as such can provide the much needed validation for a much more expensive toolkit that a quantum computer is. Since the angle encoding has consistently performed on and above par in this work, it merits further investigation. The role of entanglement, as of now, is unclear. Ref. \cite{moradi2022clinical} compared the efficiency of different encoding strategies for QML classification tasks before. They contrasted amplitude encoding against a linear in qubit number complexity encoding and found amplitude encoding to perform better. However, the underperforming encoding of theirs did have entanglement (and two qubit entangling gates). In this work, we compare amplitude encoding to an encoding (the ``angle encoding'') with no two qubit entanglng gates. Since, the encoding with no two qubit entangling gates (angle encoding) is the more consistent one, there may be benefit in QML short of using quantum correlations as well.

Although a general result was not found, as entanglement helped for some datasets, and was unhelpful for others, the performance improvements in the positive cases were quite massive. Therefore, as algorithms, the quantum algorithms remain as relevant as ever. Since, short of actual implementations on quantum hardware, classical simulations can produce the improvements for real life datasets, there is real utility in considering quantum inspired algorithms for current industry and academic problems. From a number of tests and experimentation, we can be cautiously optimistic about higher ARI performance from quantum algorithms with statistical preprocessing of data.

\section*{Acknowledgments}
The authors thank Sapienly, Inc. for providing support and resources.

\section*{Data and Code Availability}
The datasets used in this work are freely available from the following sources or open source repositories. 

%%===================================================%%
%% For presentation purpose, we have included        %%
%% \bigskip command. Please ignore this.             %%
%%===================================================%%
\bigskip
\begin{flushleft}%
Dataset source for

\bigskip\noindent
iris dataset: \url{https://archive.ics.uci.edu/dataset/53/iris}

\bigskip\noindent
wine dataset: \url{https://archive.ics.uci.edu/dataset/109/wine}

\bigskip\noindent
seeds dataset: \url{https://archive.ics.uci.edu/dataset/236/seeds}

\bigskip\noindent
glass dataset: \url{https://archive.ics.uci.edu/dataset/42/glass+identification}

\bigskip\noindent
Palmer Penguins dataset: \url{https://archive.ics.uci.edu/dataset/690/palmer+penguins-3}

\bigskip\noindent
Algerian forest fires dataset: \url{https://archive.ics.uci.edu/dataset/547/algerian+forest+fires+dataset}

\bigskip\noindent
Wholesale customers dataset: \url{https://archive.ics.uci.edu/dataset/292/wholesale+customers}

\bigskip\noindent
E Coli dataset: \url{https://archive.ics.uci.edu/dataset/39/ecoli}

\end{flushleft}

The authors should be contacted for inquiries about the code.

\begin{appendices}

%\section{Section title of first appendix}\label{secA1}

%An appendix contains supplementary information that is not an essential part of the text itself but which may be helpful in providing a more comprehensive understanding of the research problem or it is information that is too cumbersome to be included in the body of the paper.

%%=============================================%%
%% For submissions to Nature Portfolio Journals %%
%% please use the heading ``Extended Data''.   %%
%%=============================================%%

%%=============================================================%%
%% Sample for another appendix section			       %%
%%=============================================================%%

%% \section{Example of another appendix section}\label{secA2}%
%% Appendices may be used for helpful, supporting or essential material that would otherwise 
%% clutter, break up or be distracting to the text. Appendices can consist of sections, figures, 
%% tables and equations etc.

\end{appendices}

%%===========================================================================================%%
%% If you are submitting to one of the Nature Portfolio journals, using the eJP submission   %%
%% system, please include the references within the manuscript file itself. You may do this  %%
%% by copying the reference list from your .bbl file, paste it into the main manuscript .tex %%
%% file, and delete the associated \verb+\bibliography+ commands.                            %%
%%===========================================================================================%%

\bibliography{minimalist}% common bib file
%% if required, the content of .bbl file can be included here once bbl is generated
%%\input sn-article.bbl

\end{document}